# Ultraviolet-Based Science in the Solar System: Advances and Next Steps
A white paper submitted to the Planetary Science and Astrobiology Decadal Survey 2023-2032

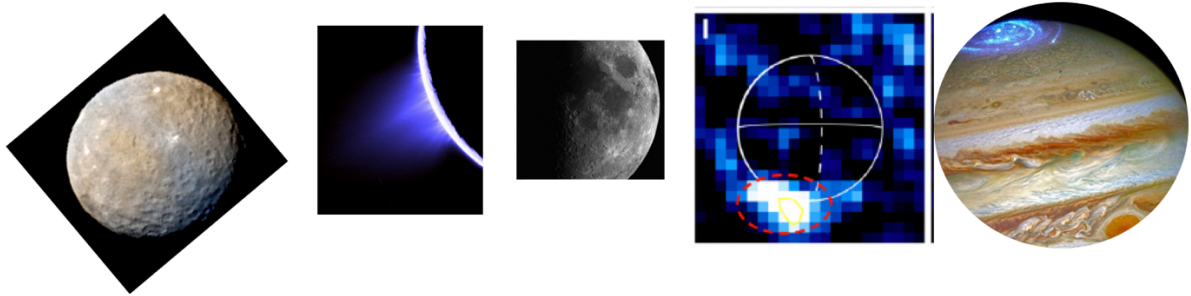

co-authors:
Amanda R. Hendrix, arh@psi.edu, Planetary Science Institute
Tracy M. Becker, SwRI
Dennis Bodewits, Auburn Univ.
E. Todd Bradley University of Central Florida
Shawn Brooks, JPL
Ben Byron, Univ. Texas San Antonio
Josh Cahill, JHU/APL
John Clarke, Boston University
Lori Feaga, U. Maryland
Paul Feldman, JHU (retired)
G. Randall Gladstone, SwRI
Candice J. Hansen, PSI
Charles Hibbitts, JHU/APL
Tommi T. Koskinen, LPL, Univ. Arizona
Lizeth Magaña, Univ. Texas San Antonio
Philippa Molyneux, SwRI
Shouleh Nikzad, JPL
John Noonan, LPL, Univ. Arizona
Wayne Pryor, Central Arizona College
Ujjwal Raut, SwRI
Kurt D. Retherford, SwRI
Lorenz Roth, KTH
Emilie Royer, PSI
Ella Sciamma-O'Brien, NASA Ames
Alan Stern, SwRI
Karen Stockstill-Cahill, JHU/APL
Faith Vilas, PSI
Bob West, JPL

co-signers:

Scott Edgington, JPL
Richard J. Cartwright, SETI Inst.
Timothy A. Livengood, U. Md
Melissa McGrath, SETI Inst.
Cesare Grava, SwRI
Joshua Kammer, SwRI
Devanshu Jha
Jian-Yang Li, PSI
Maria Gritsevich, FGI
Melissa Lane, Fibernetics LLC
Rebecca Schindhelm, Ball
Nick Schneider, LASP/CU
Greg Holsclaw, LASP/CU
Josh Colwell, UCF
Kathleen Mandt, JHU/APL

In the past decade, numerous important advances have been made in solar system science using ultraviolet (UV) spectroscopic techniques. Formerly used nearly exclusively for studies of giant planet atmospheres, planetary exospheres and cometary emissions, UV imaging spectroscopy has recently been more widely applied. The geyser-like plume at Saturn's moon Enceladus was discovered in part as a result of UV stellar occultation observations, and this technique was used to characterize the plume and jets during the entire Cassini mission. Evidence for a similar style of activity has been found at Jupiter's moon Europa using Hubble Space Telescope (HST) UV emission and absorption imaging. At other moons and small bodies throughout the solar system, UV spectroscopy has been utilized to search for activity, probe surface composition, and delineate space weathering effects; UV photometric studies have been used to uncover regolith structure. Insights from UV imaging spectroscopy of solar system surfaces have been gained largely in the last 1-2 decades, including studies of surface composition, space weathering effects (e.g. radiolytic products) and volatiles on asteroids (e.g. [2][39][48][76][84]), the Moon (e.g. [30][46][49]), comet nuclei (e.g. [85]) and icy satellites (e.g. [38][41-44][45][47][65]). The UV is sensitive to some species, minor contaminants and grain sizes often not detected in other spectral regimes.

In the coming decade, HST observations will likely come to an end. New infrastructure to bolster future UV studies is critically needed. These needs include both developmental work to help improve future UV observations and laboratory work to help interpret spacecraft data. UV instrumentation will be a critical tool on missions to a variety of targets in the coming decade, especially for the rapidly expanding application of UV reflectance investigations of atmosphereless bodies.

## 1. Significant Results from UV Studies in the last Decade

*Atmosphere and Auroral Studies.* UV observations are sensitive to small columns of gas, and as such are the key method to study planetary upper atmospheres and aurorae, i.e. the interaction of atmospheres with the space environment. UV spectroscopy has been used since the earliest space missions for atmospheric and auroral studies (e.g. [4][5][8][13][17][18][83][86]). For instance, stellar and solar occultations obtained by the Cassini Ultraviolet Imaging Spectrograph (UVIS) instrument (56-190 nm) provide unique data that probe the mesosphere and thermosphere of Titan at altitudes between 400 km and 1400 km (e.g. [14-16][55][58][61][62][79]). MAVEN/IUVS at Mars is a prime example that improved instrumentation can still result in substantial discoveries even ~50 years after the first interplanetary UV instruments (e.g. [21]).

Cassini UVIS measurements at Saturn included occultation studies of the vertical and latitudinal distribution of atmospheric constituents (e.g. [59]), atmospheric UV airglow studies [34][60], and detection of an extended planetary corona [64][80]. UVIS auroral studies at unprecedented spatial resolution (e.g. [33]) enabled detection of the Enceladus auroral footprint on Saturn [73] and many detailed multi-instrument studies of magnetospheric auroral storm development and evolution (many papers, notably [70]). UVIS spectroscopy of $H_2$ band and Lyman-alpha based auroral color ratios permitted estimation of primary auroral electron energies [35]. Reflected sunlight studies with UVIS were sensitive to the global distribution of acetylene, other hydrocarbons, and aerosols [78], allowing detection of Saturn's UV-dark north polar hexagon at stratospheric altitudes [72].

Juno has now spent 4 years in a highly elliptical polar orbit around Jupiter, and is ~75% through its nominal mission of 36 orbits. The UVS (68-210 nm) on Juno helps to make sense of the highly valuable in situ data by providing context maps, which give the particle and field investigators a sense of what is happening at other places in Jupiter's magnetosphere, far from Juno [32]. Since Juno spins once every 30 seconds, and moves at about 60 km/s when near Jupiter, building UV auroral maps requires a photon-counting, wide field-of-view, imaging spectrograph like Juno-UVS, which has a scan mirror to raster across Jupiter's enormous (and very bright) auroral regions.

***Ocean Worlds & Outer Solar System Targets.*** The advantages of UV imaging spectroscopy for detecting and investigating plumes and thin atmospheres via emissions and occultations (gas absorptions) at ocean worlds (and Io!) are clear (e.g. [36][75][77]). Observations acquired at UV wavelengths have been a critical component in studying and characterizing ocean worlds. UV measurements played a key role in the discovery and characterization of Enceladus' plume, beginning with stellar occultations observed in 2005 with Cassini's Ultraviolet Imaging Spectrograph (UVIS). These observations enabled determination of the plume's primary composition ($H_2O$ vapor), column density, and mass flux [36], along with the non-global distribution of vapor. Earth-based UV observations using HST have shown possible plume activity at Europa as well, using both emission measurements [77] and absorption measurements during Jupiter transits [81]. Cassini UVIS spectral and occultation measurements of Saturn's rings provided important constraints on structure, composition and particle size distribution (e.g. [6][11-12]). UVIS observations were coupled with HST measurements to fill Cassini's gap in spectral coverage (200-350 nm), a region of strong absorption in the outer solar system [20]51], likely the result of organics.

At Pluto, the Alice UV spectrometer (52-187 nm) aboard New Horizons observed stellar and solar occultations resulting in the only individual abundance neutral gas profiles ever obtained, for all of the major and many minor species as a function of altitude [31][56][85][91], and also set $10^4$-$10^6 \times$ stricter limits on Charon's atmosphere [93] than possible from groundbased techniques. Additionally, Alice airglow observations of Pluto's atmosphere [82] discovered the first ions high in Pluto's atmosphere ($N^+$) and detected airglow emissions from H, N, NII, $N_2$, CO and methylacetylene (a new trace species) above the disk of Pluto; argon upper limits were also set. The Alice spectrometer also detected surface reflectance in the FUV at a surprisingly high 17% between 140 and 185 nm [82].

***Icy Moon Surfaces.*** Cassini UVIS made numerous observations of the Saturnian satellites, providing clues to their surfaces and processes. In particular, UVIS measurements provided the first-ever detection of the strong $H_2O$ ice FUV absorption feature on an icy body, near 165 nm [41] and the $H_2O$ absorption has since been observed nearly ubiquitously throughout the Saturn system (e.g. [44]). Shortward of ~165 nm, $H_2O$ ice is very dark and nearly all non-$H_2O$ contaminants are brighter than $H_2O$ ice at these wavelengths. The overall dark reflectivity of the moons at wavelengths longward of 165 nm, where $H_2O$ ice is bright, can be indicative of the composition of the non-water ice contaminant. The presence of a small amount of $NH_3$ and a small amount of tholins (lab-generated materials representing organic materials) in addition to $H_2O$ ice on Enceladus' surface was suggested to explain the low FUV reflectance. Furthermore, the Cassini UVIS observations of the likely presence of photolytically-produced hydrogen peroxide ($H_2O_2$) was proposed to explain seasonal and latitudinal variations at Mimas and Tethys [47]. Another darkening agent, to which UV wavelengths are particularly sensitive, is carbon-rich organics and their weathering products. A small amount of organics sourced from Enceladus' ocean find their way into the E ring [71] and are eventually deposited onto the surfaces of the moons that orbit Saturn within the E ring. Such organics are known to become darker and redden (at UV-visible wavelengths) with weathering, and eventually spectrally flatten with carbonization, which can explain much of the coloring seen on the Saturn system surfaces [51].

***Asteroids.*** HST/STIS has observed main-belt asteroids including Ceres (115-570 nm) and (16) Psyche (170-310 nm). Ceres measurements in the far-UV confirmed the presence of a "bump" in the spectrum that is consistent with a graphitized carbon [48] and could explain Ceres' overall low albedo. HST observations of the asteroid (16) Psyche in the UV revealed an absorption band near 250 nm that is consistent with iron-oxide on the surface; the first such detection on an asteroid [7]. The steep, red UV spectral slope of Psyche is consistent with the presence of iron. Shortward of 200 nm, Psyche's spectrum becomes blue; this could indicate a spectral "bump" like that seen on Ceres and therefore diagnostic of surface composition, but more data in the far-UV are needed. HST/STIS also obtained UV spectra of (617) Patroclus and its companion Menoetius, a binary system target of the upcoming Lucy mission. The two Trojan asteroids were found to have similar mid-UV reflectance, consistent with a shared evolution as a binary rather than a scenario in which Menoetius was captured later. Small differences in the 230-260 nm region hint at possible variations in composition across the surfaces of the two bodies. The space weathering effects on asteroids were studied using HST and International Ultraviolet Explorer (IUE) data [39-40]. Rocky S-complex asteroids exhibit lunar-like space weathering effects in the UV, namely a spectral bluing with exposure due to the production of nanophase iron in the grains. Carbonaceous C-complex asteroids could experience a bluing with weathering, as a weakening of the UV absorption, which might be related to the weathering production of an opaque such as amorphous or graphitized carbon.

***Diagnostic carbon-related spectral features.*** Carbon compounds are ubiquitous in the solar system but are challenging to study using remote sensing due to the mostly bland spectral nature of these species in the traditional visible-near infrared regime. In contrast, carbonaceous species are spectrally active in the UV but have largely not been considered for studies of solar system surfaces. A compilation of existing UV data of carbon compounds -- well-studied in contemplation of the interstellar medium extinction -- was used to review trends in UV spectral behavior [50]. Thermal and/or irradiation processing of carbon species results in the loss of H and ultimately graphitization. Graphitization produces distinct spectral features in the UV (e.g. [3]). The presence or lack of such a feature at carbonaceous bodies throughout the solar system could be an important indicator of exposure age. Polycyclic aromatic hydrocarbons and fluorescence by organic materials also exhibit widely varying and diagnostic UV-visible spectral shapes (e.g. [54][63]). These characteristics must be further explored, and also exploited to better understand carbon-rich bodies throughout the solar system, in the next decade. UV-visible fluorescence is a key technique just beginning to be utilized but will be critical not only for Mars studies for surface studies in the outer solar system and asteroid belt as well.

***Comets.*** UV spectroscopy for studying cometary emissions is also well-established (e.g. [24]) and can be used to remotely monitor comet activity and to detect the presence molecules that have weak or difficult-to-observe emission features (e.g. $S_2$; H and O for $H_2O$; $O_2$ through OI; $CO_2$ through CO and $CO_2^+$ (e.g. [9][19][28][89][90]). The Rosetta Alice FUV (70-205 nm) instrument made both surface and coma measurements at comet 67P/C-G, resulting in a low geometric albedo of ~5% and a surprising blue-sloped surface spectra below 200 nm compared to its red-sloped surface at NUV and visible wavelengths [22]. The FUV coma measurements include signatures of every dominant volatile found in comets, $H_2O$ (directly detected in absorption of background stellar continuum [57], CO (directly detected via the Fourth Positive fluorescence [28], $CO_2$ (inferred by the CO Cameron bands [28], and $O_2$ (directly detected in absorption of background stellar continuum [57], as well as dissociative products of the aforementioned molecules, H, O, and C (directly detected in emission from fluorescence and electron impact [26-28][69]. Both the parent and product species can be measured simultaneously, leading to more robust results. Several of the Rosetta results were unexpected and even paradigm changing, like the high $O_2/H_2O$ abundances in a comet (from 5% to >50% [57], outbursts that were dominated uniquely by dust or gas, but not both, or expelled ice [1][27], and the interaction of the coma with a passing CME event showing the importance of including electron impact in coma physics [69]. These measurements improve our understanding of cometary composition, and thus inferences about the conditions in the solar nebula from which comets formed, the mechanism and volatiles responsible for driving outbursts, the physics in the coma causing the dissociation and excitation of volatile species, and the lifetime of volatile species in the coma environment once emitted by the comet, which gives important context for the secondary species more readily measured from the ground and how their abundances relate to that of the parent volatiles directly released from the nucleus.

***The Moon***. The lunar exosphere was studied in the UV in the Apollo 17 mission [23] and study continues with the Lyman Alpha Mapping Project (LAMP) spectrograph (57-196 nm) on board the Lunar Reconnaissance Orbiter (LRO) (e.g. [25]). For surface studies of the Moon, LRO/LAMP implements a novel technique of utilizing UV-bright stars and the interplanetary medium (IPM) Lyman-$\alpha$ glow to observe permanently shadowed regions, some of the coldest regions in the solar system where ice deposits remain cold-trapped. LAMP observations have served to constrain the $H_2O$ ice abundance of these deposits, <2% by mass [30][37]. More than a decade after launch, current LAMP analyses are further characterizing condensed $H_2O$ deposits, while exploring the possibility of ultra-cold $H_2O$ ice populations and the possible existence of additional condensed volatile species within PSRs. Dayside far-UV maps are also obtained using the more traditional photometry technique with the Sun as the illumination source, and are used to sense migrating lunar $H_2O$ [46][52] and composition (e.g. [10]). Together, these LRO-LAMP measurements provide a unique perspective on the lunar "hydrological cycle," connecting the surface abundance of $H_2O$ frost trapped in the Moon's cryosphere to volatile transport processes involving the lunar exosphere.

## 2. Support for Development of UV Systems is Needed

Approaching the next decade, advancements in UV-related technologies (detectors, gratings, electronics miniaturization) are needed to advance to the next steps in planetary science. Weak signals at outer solar system targets (e.g. KBOs, Trojan asteroids, moons of Uranus and Neptune), for instance, will require utilization of more sensitive and high dynamic range detectors to fully take advantage of the UV-diagnostic spectral clues. Additionally, orbital missions are not the only place for UV instrumentation – landers and rovers can also benefit from this technology, for *in situ* studies, as can small satellites piggy-backing on larger missions.

Discoveries of Rosetta, Cassini, and HST instruments have further shown that observations in the UV, an area rich with spectroscopic diagnostic lines, provide key information in planetary bodies both in remote and in situ formats (e.g., UV Raman spectroscopy that is part of the Mars 2020 instrument suite). What has further been shown from these successful observations and near-future exploration is that making new discoveries beyond current missions requires advancements in technology and instrumentation. While much has been gleaned with previous observations, the need to advance detector technology development has been shown, as has the need for development of compact and powerful spectrometers, cameras enabled by more advanced detectors, gratings, better optical coatings, and low power electronics with small footprint.

In detector area, where often the tallest tent poles lie, a solid-state detector that does not require high voltage or vacuum tube technology for operation enables more compact instruments. Solar-blind, high efficiency UV detectors—some with single photon sensitivity, radiation hardness, high dynamic range and high operating temperatures will be transformational for UV instruments.

Fortunately, at the same time there has been demonstration of potentially disruptive technology that can be capitalized to create the next generation of powerful, robust and affordable instruments. Some of the major areas of innovation that will be of use in the coming decade are: 2D doping of silicon detectors (including delta doping and superlattice doping); Quanta Image Sensors (QIS) [29] without impact-ionization gain and therefore without the associated dark current penalty; electron multiplying CCDs (EMCCDs), detector-integrated filters [53][66][68], and potentially metamaterials as a way to produce high operating temperature, solid state, silicon-based high efficiency UV detectors with solar blind capability.

Another area of solar blind, photon counting detectors - both in solid-state and photoemissive form - is gallium nitride. Photocathodes with high efficiency and air-stable capability [87] as well as APDs [67] have been demonstrated. Their wide bandgap, direct bandgap, and wide range due to alloying AlN and InN, mean that this family of material offers inherent visible blindness, higher operating temperature, and potentially radiation tolerance.

### 3. UV Laboratory Work is Needed

Laboratory studies of planetary analogs are critical for interpretation of remote sensing UV data. Optical properties of materials common to planetary surfaces are yet to be characterized in the UV. Optical constants of condensed volatiles, regolith simulants and returned samples are sorely lacking, compared to infrared counterparts. Because UV investigations are sensitive to composition of the top most (~ 100 nm depending on composition) layer of regoliths, these wavelengths are sensitive to effects of plasma-surface interactions, especially for planetary bodies immersed in a magnetosphere. UV lab studies (e.g. reflectance spectra of candidate species and mixtures) are critically needed to support and interpret the acquired spacecraft data.

Over the last several decades, it has been incredibly difficult to begin to obtain UV measurements using NASA R&A funds. However, we are pleased to report that, after numerous unsuccessful attempts in traditional R&A programs, UV capabilities at several labs have been developed in recent years, including at SwRI, PSI, LASP and JHU-APL. At PSI and LASP, the majority of the support to develop the labs and take initial data came from SSERVI, along with internal funding. All of the support for the SwRI and APL labs came from internal funding. Great work is being accomplished in these labs but **NASA R&A support is needed to make further essential measurements**. UV lab studies are notoriously tricky, owing to stringent contamination control requirements. The measurements are both difficult to make and are time-consuming. Often UV spectra are obtained by using a monochromer, in order to keep scattered light to a minimum, and can take hours to collect a single useable spectrum. It is important that upkeep of instrumentation and measurements of different types in these labs continue; avenues for funding of these activities in NASA R&A programs would be incredibly useful. Adopting a NASA-sponsored model equivalent of RELAB to establish a multi-user spectroscopy facility to advance ultraviolet laboratory investigations is urgently needed. We urge the Decadal Survey to support continued funding, via R&A programs, of UV laboratory work.

Both qualitative (i.e. reflectance spectra) and quantitative measurements (i.e. derived optical constants) are needed from the far-UV (~100 nm) into the near-UV (~400 nm). Materials include: amorphous ice and crystalline ice (as a function of temperature), hydrated salts and acids (pristine and weathered), $SO_2$ ice, $CO_2$ ice, many organics (including analogs of atmospheric aerosols which could settle on the surface of some bodies), and radiation products both of 'weathered' silicates and of 'weathered' ices with the appropriate radiation and at appropriate temperature. Besides reflectance and transmission measurements for the derivation of optical constants, electron impact measurements and the spectral characterization of small molecules and their ions are needed. Many electron impact emission features (observed by Rosetta Alice and OSIRIS at 67P) have only been measured at fixed impact energy, but not between their threshold energy and ~100 eV, the energy range that includes the majority of electrons observed at the comet. Furthermore, currently available laboratory measurements of absorption cross sections are generally acquired at room temperature. Measurements to characterize the temperature dependence of absorption cross-sections are needed (e.g. for $H_2O$ with applications for Enceladus). Of importance is proper archiving of the laboratory data, enabling community access.

## 4. Open Questions and UV Needs for Future Missions

Though tremendous accomplishments have been made using UV instrumentation in the last decades (as outlined in Sec 1), there is still much to do. Future missions should carry modern UV instrumentation to continue discovering the characteristics of solar system targets. UV instrumentation needs to be included in future Discovery and New Frontiers class missions, on lander missions as well as on orbiter/flyby missions. For future spacecraft payloads, the typical gap in spectral coverage in the 200-1000 nm region must be removed - this is a spectrally active region where data are needed and useful for diagnostics. Just as Cassini and HST UV measurements have played important roles in characterizing plume activity on Enceladus and Europa, upcoming measurements by the UVS onboard Europa Clipper will be critical in finding and characterizing plumes on Europa. Upcoming missions to potential ocean worlds such as Triton should include UV instrumentation in their payloads to search for activity.

Currently, Hubble time for solar system targets is minimal (over the entire life of the mission 6% of the total science time goes to solar system targets), and though future potential astrophysics platforms (if they include UV capability) such as HabEx or LUVOIR will be useful to planetary science, a UV-capable mission *dedicated to solar system targets* will be extremely valuable in the post-Hubble era. [92]. A dedicated solar system telescope has different requirements than typical astrophysical facility, including tracking capabilities, large FOV, rapid response capability, and the ability to handle bright extended sources.

**The Search for Water and OH on Asteroids and the Moon:** The far-UV is extremely sensitive to the presence of $H_2O$ and does not require any additional thermal corrections. Even trace amounts of $H_2O$ induce a sharp increase in reflectivity at ~165 nm [74] which has been detected in lunar soil from the LRO LAMP observations [52]. Future far-UV observations of asteroids should be conducted to search for $H_2O$, especially since near-IR measurements can be difficult to interpret due to the proximity of the OH band to the 3-μm water feature in addition to the required thermal corrections in the IR. Furthermore, future landed or orbital UV measurements at the Moon can better characterize $H_2O$ and OH (via the 308 nm emission) to better constrain the hydrological cycle at the Moon.

**Planetary aurorae** are best studied at UV wavelengths, where the bulk of the emission is produced and the level of reflected sunlight is low, i.e., the highest contrast that can be obtained when observing the sunlit face of a planet. Remote observing campaigns with HST have been carried out in parallel with *in situ* measurements of the local plasma environment during the Galileo, Cassini, and Juno missions. A future mission to an Ice Giant planet will similarly depend on remote observations to establish the global picture of auroral activity.

## 5. Conclusions and Summary.

In the next decade, UV measurements will play critical roles in understanding ocean worlds as well as small bodies, the building blocks of our solar system. Facilities such as LRO/LAMP and Cassini UVIS have demonstrated how UV reflectance maps constrain surface composition in new, previously unappreciated ways. The capabilities of UV systems for studying ices, gases and organics are particularly critical. Guidance from the decadal survey is needed to sustain and promote these important measurements and their supporting infrastructure, including an HST follow-on facility dedicated to solar system targets with UV capability, UV instrumentation on deep-space missions, laboratory measurements and UV instrument development.